\documentstyle[12pt,aaspp4]{article}
\def \yskip{\penalty-50\vskip3pt plus 3pt minus 2pt}
\def \pp{\par \yskip \noindent \hangindent .4in \hangafter 1}
\def \abc#1#2#3#4 {\pp#1, {\sl#2}, {\bf#3}, #4}
\def \blank {\lower 5pt\hbox to 0.75in{\hrulefill}}
\def\etal{{\it et al.}}

\newfont{\rten}{cmr10}

\begin{document}

\vspace*{-0.5cm}
\normalsize
 
\title{Supernova Limits on the Cosmic Equation of State}
\vspace*{0.3cm}

Peter M. Garnavich\footnote{Harvard-Smithsonian Center for Astrophysics, 60 Garden St., 
Cambridge, MA 02138},
Saurabh Jha$^1$,
Peter Challis$^1$,
Alejandro Clocchiatti\footnote{Departmento de Astronom\'ia y Astrophisica Pontificia
Universidad Cat\'olica, Casilla 104, Santiago 22, Chile},
Alan Diercks\footnote{Department of Astronomy, University of Washington, Seattle, WA 98
195},
Alexei V. Filippenko\footnote{Department of Astronomy, University of California, 
Berkeley, CA 94720-3411},
Ron L. Gilliland\footnote{Space Telescope Science Institute, 3700 San Martin Drive, Baltimore, MD 21218},
Craig J. Hogan$^3$,
Robert P. Kirshner$^1$,
Bruno Leibundgut\footnote{European Southern Observatory, Karl-Schwarzschild-Strasse 2,  
Garching, Germany},
M. M. Phillips\footnote{Cerro Tololo Inter-American Observatory, Casilla 603, 
La Serena, Chile},
David Reiss$^{3}$,
Adam G. Riess$^4$,
Brian P. Schmidt\footnote{Mount Stromlo and Siding Spring Observatory, Private Bag, 
Weston Creek P.O.,  Australia},
Robert A. Schommer$^7$,
R. Chris Smith\footnote{University of Michigan, Department of Astronomy, 
834 Dennison, Ann Arbor, MI 48109},
Jason Spyromilio$^{6}$,
Chris Stubbs$^{3}$,
Nicholas B. Suntzeff$^7$,
John Tonry\footnote{Institute for Astronomy, University of Hawaii, Manoa, HI 96822},
Sean M. Carroll\footnote{Institute for Theoretical Physics, University of California, Santa
Barbara, CA 93106}


\begin{abstract}

We use Type~Ia supernovae studied by the High-Z Supernova Search Team to
constrain the properties of an energy component which may have contributed to
accelerating the cosmic expansion. We find that for a flat geometry
the equation of state parameter for the unknown component,
$\alpha_{\rm x}=P_{\rm x}/\rho_{\rm x}$, must be less than $-0.55$ (95\% confidence) for any value
of $\Omega_{\rm m}$ and is further limited to $\alpha_{\rm x} <-0.60$ (95\%)  if
$\Omega_{\rm m}$ is assumed to be greater than 0.1 . These values are inconsistent 
with the unknown component being topological defects such as domain walls,
strings, or textures. The supernova data are consistent with a cosmological 
constant ($\alpha_{\rm x}=-1$) or a scalar field which has had, on average, an equation of state 
parameter similar to the cosmological constant value of $-1$ over
the redshift range of $z\approx 1$ to the present. Supernova and
cosmic microwave background observations give complementary
constraints on the densities of matter and the unknown component.
If only matter and vacuum energy are considered, then the current
combined data sets provide direct evidence for a spatially flat
Universe with $\Omega_{\rm tot}=\Omega_{\rm m}+\Omega_\Lambda = 0.94 \pm 0.26$ ($1\sigma$).

\end{abstract}


\section{Introduction}

Matter that clusters on the scale of galaxies or galaxy clusters is 
insufficient to close the Universe, with conventional values near
$\Omega_{\rm m} = 0.2 \pm 0.1$ (Gott \etal\ 1974; Carlberg \etal\ 1996; Lin \etal\ 1996;
Bahcall, Fan, \& Cen 1997). Observations
of distant supernovae provide credible evidence that the
deceleration rate of the Universal expansion is
small, implying that the total matter density, clustered or smooth, is insufficient
to create a flat geometry (Garnavich \etal\ 1998; Perlmutter \etal\ 1998).
Either the Universe has an open geometry or, if flat, other
forms of energy are more important than matter.

Large samples of supernovae analyzed by the
High-Z Supernova Search collaboration
(Riess \etal\ 1998a, hereinafter [Riess98]) and the Supernova Cosmology Project
(Kim 1998) now suggest that the Universe
may well be accelerating. Matter alone cannot accelerate the
expansion; so if taken at face value, the observations demand 
an additional energy component for the Universe.  While the vigorous
pursuit of possible systematic effects 
(e.g. H\"oflich, Wheeler, \& Thielemann 1998) will be important in 
understanding these observations, it is instructive to see what they imply
about the energy content of the Universe.

The cosmological constant was revived
to fill the gap between the observed
mass density and the theoretical preference for a flat Universe (Turner, Steigman, \& Krauss
1984; Peebles 1984) and also to alleviate the
embarrassment
of a young Universe with older stars (Carroll, Press, \& Turner 1992). 
The cosmological
constant is a negative pressure component arising
from non-zero vacuum energy (Weinberg 1989). It would be extraordinarily
difficult to detect on a small scale, but $\Omega_\Lambda= 1 - \Omega_{\rm m}$
could make up the difference between the matter density $\Omega_{\rm m}$ 
and a flat geometry and might be detected by measurements on a
cosmological scale. There are few 
independent observational
constraints on the cosmological constant, but Falco, Kochanek, \& Mu\~noz
(1998)
estimated that $\Omega_\Lambda < 0.7 $ (95\%\ confidence)
from the current statistics of strong
gravitational lenses. If the matter density is less than $\Omega_{\rm m}\sim 0.3$, 
this limit is
close to preventing the cosmological constant from
making a flat geometry. 
Further, a cosmological constant which just happens to be of the same order 
as the matter content at the present epoch raises the issue of ``fine tuning''
(Coles \& Ellis 1997). 
A number of exotic forms of matter which might contribute to
cosmic acceleration are physically possible and viable alternatives
to the cosmological constant
(Frieman \& Waga 1998; Caldwell, Dave, \& Steinhardt 1998).
The range of possibilities can be narrowed using supernovae
because the luminosity distance
not only depends on the present densities of the 
various energy components
but also on their equations of state while the photons we see
were in flight. 
Here, with some simplifying assumptions, we consider the
constraints that recent supernova observations place on
the properties of an energy component accelerating
the cosmic expansion.

\section{Observations}

The type~Ia supernovae (SNIa) have been analyzed 
by the High-Z Supernova Search Team
and described by Riess98, 
Garnavich \etal\ (1998), Schmidt \etal\ (1998), and Riess \etal\ (1998b). 
The full sample from Riess98 consists
of 50 SNIa. Of these, 34 are
at $z<0.2$ while the remaining 16 cover a range in redshift of
$0.3<z<1.0$. Six of the high-redshift events were analyzed using the
``snapshot'' method developed by Riess \etal\ (1998b).
This innovative technique uses high-quality spectra to deduce
information unavailable due to a poorly sampled light curve. While the
errors estimated from the snapshot method are larger than
those from direct light curve fitting, the snapshot sample
provides a significant, independent set of SNIa distances.

As shown by Phillips (1993), the light curve decline rate of
SNIa is correlated with the luminosity at maximum 
brightness of these exploding white dwarfs.
This correlation has been calibrated by Hamuy \etal\ (1996, the $\Delta m_{15}(B)$ method)
and by Riess, Press, \&\ Kirshner (1995, 1996, the Multi-Color Light Curve Shape or
MLCS method which includes  a correction for extinction), and both show that applying
this correction to the
SNIa Hubble diagram significantly reduces the scatter. 
Phillips \etal\ (1998) extended the $\Delta m_{15}(B)$ approach
to include an estimate of the extinction. In Riess98, an improved version of the 
MLCS method is presented.  Here, as in Riess98, we apply both MLCS and $\Delta m_{15}(B)$
(with extinction correction) techniques to the analysis to
gauge the systematic errors introduced by different light curve fitting methods.

\section{Analysis}

The apparent brightness of a SNIa corrected for light curve decline
rate and extinction provides an estimate of the luminosity
distance, $D_L$, from the K-corrected observed magnitude, 
$ m=M+5{\rm log}D_L+25$,
and the absolute magnitude, $M$, of SNIa.
As described by Schmidt \etal\ (1998) and Carroll, Press, \& Turner (1992),
the luminosity distance depends on the content
and geometry of the Universe in a Friedmann-Robertson-Walker cosmology  
\begin{equation}
D_L = {{c(1+z)}\over{H_0\sqrt{|\Omega_k|}}}sinn\left\{\sqrt{|\Omega_k|}
\int_0^z[\sum_{i}\Omega_i(1+z')^{3(1+\alpha_i)}+\Omega_k (1+z')^2]^{-{1}/{2}}
dz' \right\}
\end{equation}

\begin{equation}
sinn(x)=\cases{
sinh(x),&if $\Omega_k >0$;  \cr
x,&if $\Omega_k =0$; \cr
sin(x),&if $\Omega_k <0$,  \cr}
\end{equation}

\noindent where $\Omega_i$ are the normalized densities of the
various energy components of the Universe and $\Omega_k=1-\sum_{i}\Omega_i$ 
describes the effects of curvature. The exponent $n={3(1+\alpha)}$
defines the way each component density varies as the Universe expands,
$\rho\propto a^{-n}$, where $a$ is the cosmic scale factor. 
For example, $n$ has the value 3 for normal matter since the
mass density declines proportionally to the volume.
Alternatively, $\alpha_i$ is the equation of state parameter
for component $i$ defined as the ratio of the pressure to the energy
density, $\alpha_i =P_i/\rho_i$ (sometimes denoted in the literature as $w$).
The relation $n={3(1+\alpha)}$ is easily derived from the conservation of energy equation 
in comoving coordinates (e.g. Weinberg (1972) equation 15.1.21).
In the most general case, the equation of state can vary with time in ways
other than assumed here (as the sum of power laws in $1+z$), but we
are limited by the quality and range of the
supernova observations to consider only its average effect between the
present and $z<1$. 
The present-day value of the Hubble constant ($H_0$) and the absolute magnitude
of SNIa ($M$) are primarily set by the low-redshift sample, which allows the
high-redshift events to constrain the cosmological effects. This means
that conclusions derived from SNIa are independent of the absolute distance scale.

Gravitational lensing by matter distributed between the observer
and supernovae at high redshift can affect the observed brightness
of SNIa and induce errors in the estimate of their luminosity
distances (Kantowski, Vaughan, \& Branch 1995). For realistic
models of the matter distribution and $\Omega_{\rm m} < 0.5$, the
most likely effect of the lensing is to make the supernovae at $z=0.5$
about 2\% fainter  than they would appear if the 
matter were distributed uniformly (filled-beam) as shown
by Wambsganss {\it et al.} (1997). Holz \& Wald (1997) have
shown that the magnitude of the effect also depends on
whether the matter is distributed smoothly on galaxy scales or is
clumped in MACHOS,
but the error induced remains small when $\Omega_{\rm m} < 0.5$.
For simplicity, our calculations consider only
the filled-beam case, however, the effect of assuming
the extreme case of an empty-beam is shown by Holz (1998). 

There are a few known, and possibly some unknown, energy components that 
affect $D_L$. 
Ordinary gravitating matter, $\Omega_{\rm m}$, certainly has had some
effect on the Universal expansion between $z\approx 1$ and now. Since the
matter density scales inversely with the volume, $\alpha_m=0$,
and matter (baryons, neutrinos, and dark matter; formerly Earth, Air, and Water)
contributes no pressure. 
Radiation (Fire in an earlier lexicon) ($\alpha_{\rm r}=+1/3$) dominated
during a period in the early Universe but is negligible for $z<1$.
Equation~1 shows that for non-flat models the curvature term, $\Omega_k$,
contributes to the luminosity distance like a component
with $\alpha_k=-1/3$, but additional geometrical effects as prescribed by
equation~2 are also important.

Other more speculative components have been proposed. A non-zero vacuum energy,
$\Omega_\Lambda$, is a popular possibility explored by Riess98 for
this data set. Because the vacuum energy
density remains constant as the Universe expands (that is, 
$\rho_\Lambda\propto a^0$), we have $\alpha_\Lambda=-1$.  Topological
defects created in the early Universe could also leave remnants that
might contribute to the energy now. Networks of cosmic strings may be
a natural consequence of phase transitions in the young Universe
and if they did not intercommute would have an average effective $\alpha_{\rm s}=-1/3$
(Vilenkin 1984; Spergel \& Pen 1997). A network of comoving
domain walls would have an average equation of state
parameter of $-2/3$ (Vilenkin 1985) while a globally wound texture would
produce an $\alpha_{\rm t}={-1/3}$ (Davis 1987; Kamionkowski \& Toumbas 1996). 

Evolving cosmic scalar fields with suitable potentials
could produce a variety of exotic equations of state
with significant densities at the present epoch 
(Peebles \& Ratra 1988; Frieman \etal\ 1995; Frieman \& Waga 1998). 
Scalar fields could also produce variable mass particles
(VAMPS) which would redshift more slowly than ordinary matter
creating an effective $\alpha_{\rm VAMP} <0$ (Anderson \& Carroll 1998).
These fields may evolve over time and would produce an interesting
variety of cosmic histories.  Our goal is modest:
we only hope to constrain the average $\alpha$ over the
range where SNIa are presently observed.

To simplify the analysis, we assume that
only one component affects the cosmic expansion in addition to 
gravitating matter.
Because the origin of the acceleration is unknown, we will refer to this
as the ``X'' component with a density of $\Omega_{\rm x}$ and equation of
state of $P_{\rm x} =\alpha_{\rm x}\rho_{\rm x}$. Caldwell, Dave, \& Steinhardt (1998) have
dubbed the unknown component ``quintessence'' as the other four essences
have already been employed above.
We assume that the Universe on very large scales is
accurately described by general relativity and that the
``X'' component obeys the
null energy condition (NEC). The NEC states that, for any
null vector $v^\mu$, the energy-momentum tensor satisfies $T_{\mu\nu}v^\mu v^\nu \geq 0$
(see, e.g., Wald 1984). This is the weakest of all conventional energy
conditions, and should be satisfied by any classical
source of energy and momentum including those discussed above. 
In a Robertson-Walker  metric, the NEC is
equivalent to requiring $\rho_{\rm x} + P_{\rm x} \geq 0$.
The NEC therefore restricts the energy density of the unknown component
to be positive for $\alpha_{\rm x} >-1$ and negative when $\alpha_{\rm x} <-1$
while the energy density of the cosmological constant ($\alpha_{\rm x} =-1$) is unconstrained.

\clearpage
\begin{figure}[h]
\plotfiddle{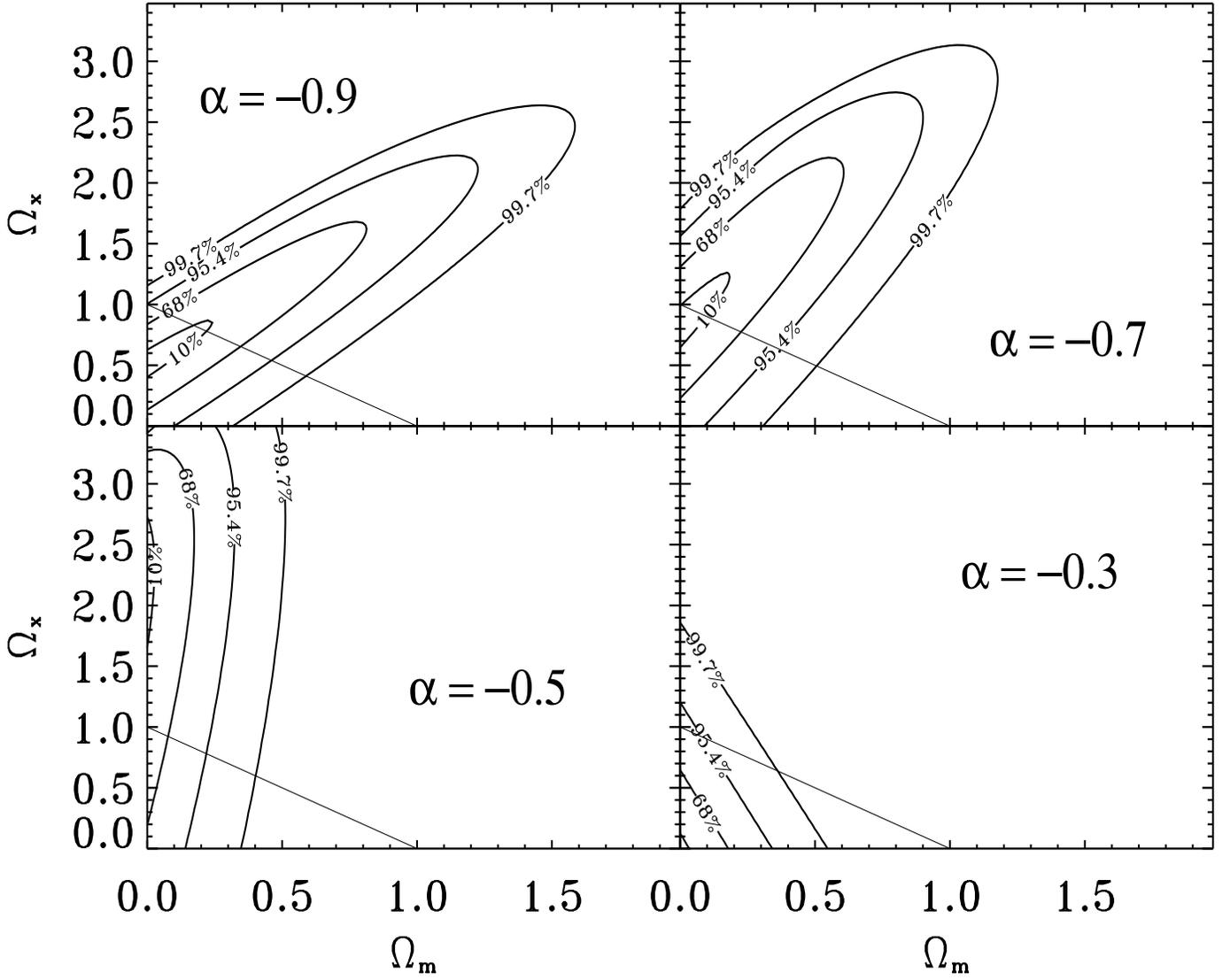}{10.0cm}{-90}{90.0}{100.0}{-350.0}{390.0}
\vspace*{3.5cm}
\caption{The joint probability distributions for $\Omega_{\rm m}$
and the density of the unknown component, $\Omega_{\rm x}$, based on the
SNIa magnitudes reduced with the MLCS method. Four representative values
of the equation of state parameter, $\alpha_{\rm x}$, are shown. See
Riess98 for the distribution when $\alpha_{\rm x} = -1$.}
\end{figure}

\section{Results}

First, we fix the equation of state of the unknown component and estimate
the probability density function for the parameters
$\Omega_{\rm x}$, $\Omega_{\rm m}$, and $H_0$ given the observed SNIa distance moduli.
The joint likelihood distributions
are then calculated in the same way as by Riess98 and shown for representative
values of $\alpha_{\rm x}$ in Figure~1. Here we integrate over all possible $H_0$
with the prior assumption that all values are equally likely.
For $\alpha_{\rm x}<-0.7$, the
derived constraints are similar to those found by Riess98 for a cosmological constant
($\alpha_{\rm x}=-1$). However, as the equation of state parameter increases, 
the major
axis of the uncertainty ellipses rotates about a point on the $\Omega_{\rm x}=0$
line. For an accelerating Universe, the pivot point is on the negative
$\Omega_{\rm m}$ side. When $\alpha_{\rm x}>-0.4$, the ``X'' component could not reproduce
the observed acceleration and all of these models give a poor fit to the observed
SNIa data: the best fit occurs for a completely empty Universe.

Next, we allow the equation of state to vary freely but
restrict the densities to $\Omega_{\rm m}+\Omega_{\rm x}<1$, or open models.
We then integrate over all possible values of $\Omega_{\rm x}$ assuming
a uniform prior distribution to provide the joint probability for
$\alpha_{\rm x}$ and the matter density. From the NEC we must include
regions where $\alpha_{\rm x} <-1$ and $\Omega_{\rm x}$ is negative, but
these are unable to produce an accelerating Universe so they have a very low
probability. For open models, highest joint probabilities are confined to
a region bounded by $-1.0< \alpha_{\rm x} < -0.4$ and $\Omega_{\rm m} <0.2$.
If we consider any value of $\Omega_{\rm m}$ equally likely, then
$\alpha_{\rm x} < -0.47$ for the MLCS method and $\alpha_{\rm x} < -0.64$
for the $\Delta m_{15}(B)$ results with 95\%\ confidence.

\begin{figure}[h]
\plotfiddle{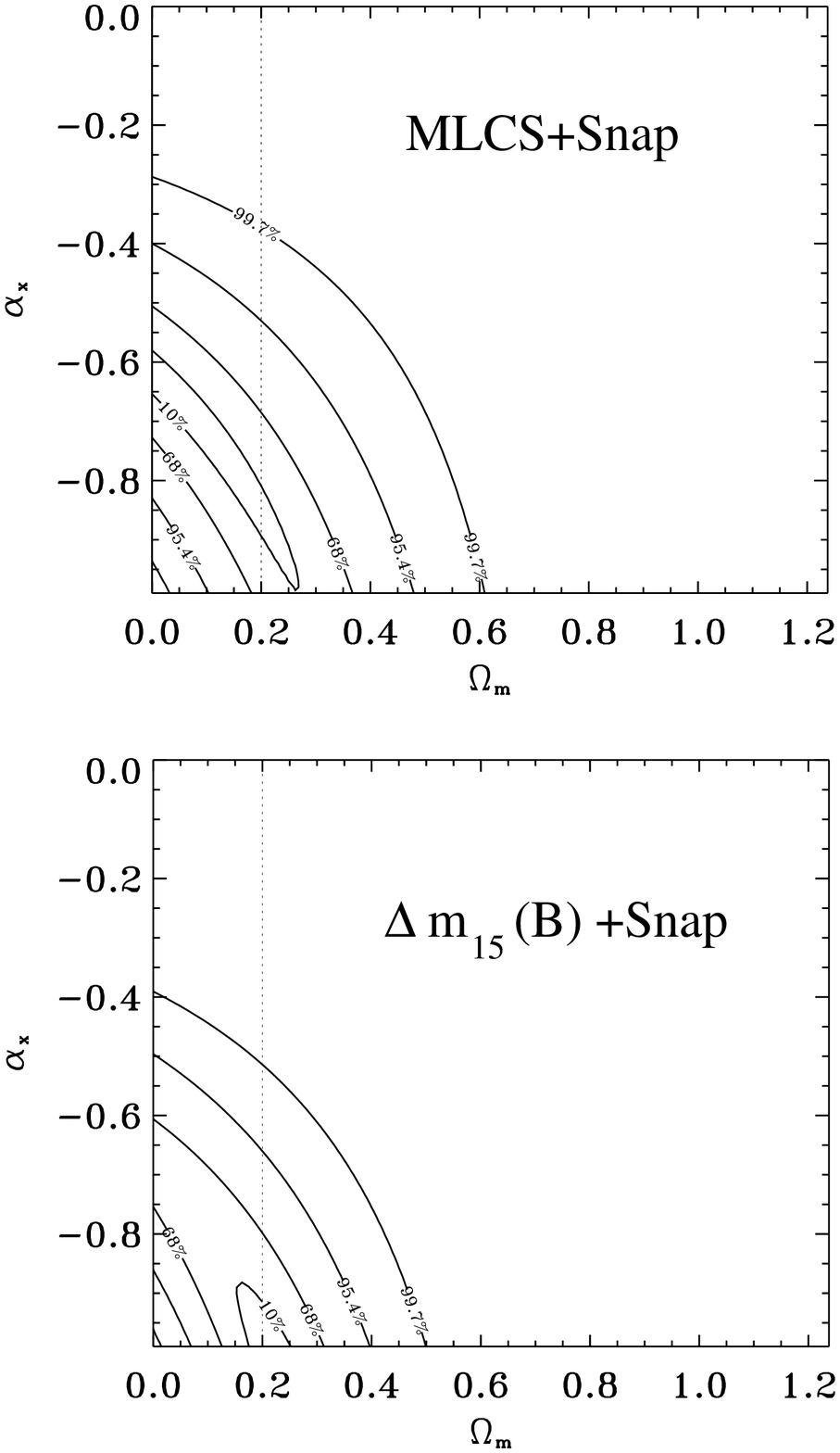}{11.0cm}{0}{80.0}{80.0}{-250.0}{-260.0}
\vspace*{8.0cm}
\caption{The joint probability distributions from
SNIa for $\Omega_{\rm m}$
and the equation of state parameter, $\alpha_{\rm x}$, assuming
a flat spatial geometry ($\Omega_{\rm m} +\Omega_{\rm x}=1$). The top panel
uses supernova distances from the MLCS method combined with supernovae
reduced using the snapshot method, while the bottom panel
is from the $\Delta m_{15}(B)$ technique plus snapshot results. The
vertical broken line marks the matter density estimated from galaxy
cluster dynamics.  }
\end{figure}

Finally, we consider flat models for the Universe.
The joint probability between the equation of state parameter and
the matter density for $\Omega_{\rm m}+\Omega_{\rm x}=1$ is shown in Figure~2.
The two cases are for
the MLCS and the $\Delta m_{15}(B)$ light curve fits and demonstrate
that the two methods for deriving luminosity from light curves provide consistent constraints. 
Note that in the flat case,
the NEC allows $\alpha_{\rm x} <-1$ only when $\Omega_{\rm m} >1$ which has an
insignificant probability and is not plotted.  
These plots can be compared to pioneering calculations by Turner \& White (1997)
and White (1998) which used smaller supernova samples. 
The improved SNIa data favor acceleration and support both
a low $\Omega_{\rm m}$ and a small value of $\alpha_{\rm x}$. 
Integrating the probability over all values of $\Omega_{\rm m}$ assuming a
uniform prior
shows that $\alpha_{\rm x} <-0.55$ for MLCS and  $\alpha_{\rm x} <-0.63$ for $\Delta m_{15}(B)$
(95\%\ confidence). If we assume $\Omega_{\rm m} > 0.1$ then the limits
tighten to $\alpha_{\rm x} <-0.60$ (MLCS) and $\alpha_{\rm x} <-0.69$ ($\Delta m_{15}(B)$).
For matter densities near $\sim 0.2$ favored by galaxy cluster dispersions, the
most probable equation of state parameters are between $-0.7$ and $-1.0$.
These results disfavor topological defect models
such as domain walls (90\%\ confidence) and eliminate strings
and textures (99\%\ level) as the principal component
of the unknown energy.  The cosmological constant, or a form of quintessence that
resembles it for $ z < 1 $, is supported by the data. 
Constraints that refer to higher redshift are needed to narrow the 
range of possible models.

\section{Other Constraints}

High-Z SNIa observations combined with the cosmic microwave
background (CMB) anisotropy angular power spectrum provide complementary
constraints on the densities of matter and the ``X'' component
(White 1998; Tegmark {\it et al.} 1998). Details of the 
CMB power spectrum depend on a large number of
variables, but the angular scale of the first acoustic peak depends primarily
on the physics of recombination and the angular diameter distance
to the surface of last scattering (White \& Scott 1996; White 1998). 
Rather than fit the power spectrum in
detail, we
have restricted our attention to the location of the first acoustic
peak as estimated from current CMB experiments
(Hancock \etal\ 1998).  This is a rapidly moving experimental field,
and new results will surely supersede these, but they illustrate the
power of combining the supernova data with the CMB.
We employ the analytic approximations of White (1998) to
determine the wavenumber of the acoustic peak at recombination, and
those of Hu \& Sugiyama (1996) to determine the recombination redshift;
thus we assume adiabatic fluctuations generate the anisotropy. In
addition, we have ignored reionization and fixed the number of neutrino
species at three, as well as assuming only scalar modes, with a
spectral index $n = 1$. A thorough treatment of this problem would allow
all of these parameters to vary and integrate the probability over all
possible values (that is, marginalize over them); however,
this would be very time-consuming, even with the fast CMB code of
Seljak \& Zaldarriaga (1996), and disproportionate to the precision 
of the current data. A large exploration of the parameter
space involved (though lacking a full variation of $\Omega_\Lambda$)
can be found in Bartlett \etal\ (1998) and Lineweaver (1998). 

Our calculation determines the angular scale multipole of the first
acoustic peak for a grid in a three-dimensional parameter space of
($\Omega_M, \Omega_\Lambda, H_0$), where we explicitly allow for
open, flat, and closed universes with and without a cosmological
constant.  We also employ the additional constraint on the baryon
density $\Omega_b h^2 = 0.024$ ($h = H_0 / 100 \; \rm{km \, s^{-1} \,
Mpc^{-1}}$) derived from the primordial deuterium abundance and
nucleosynthesis (Tytler, Fan, \& Burles 1996). Other estimates of the
baryon fraction (see Fugikita, Hogan, \& Peebles 1998) could be used, but the
location of the peak depends only weakly on this parameter. 
Where possible, we checked these
calculations with numerical integrations (Seljak \& Zaldarriaga 1996)
and confirmed that the peak locations agree to $\lesssim 10\% $, which is
adequate for this exploration.

Following White (1998), we combine the predicted peak location
with the observations using a 
phenomenological model for the peak (Scott, Silk, \& White 1995).
Recent CMB measurements analyzed by Hancock \etal\ (1998) give the
conditional likelihood of the
first acoustic peak position as $l_{\rm peak} = 263^{+139}_{-94}$,
based on best-fit values of the peak amplitude
and low multipole normalization. Rocha \etal\ (1998) have provided
us with a probability distribution function for the first peak
position based on marginalizations over the amplitude and normalization
which is a more general approach than by Hancock \etal . The Rocha \etal\
function gives $l_{\rm peak} = 284^{+191}_{-84}$ which is only a small
shift from the value derived using the conditional likelihood method.
We then marginalize the likelihood in our three-dimensional parameter 
space over $H_0$ with a Gaussian prior based on our own SNIa result
including our estimate of the systematic error from the Cepheid distance scale,
$H_0 = 65 \pm 7 \; {\rm km \, s^{-1} \, Mpc^{-1}}$ (Riess98). 
It is important to note that the SNIa constraints on
($\Omega_{\rm m}, \Omega_\Lambda$) are independent of the distance scale,
but the CMB constraints are not. We then combine marginalized
likelihood functions of the CMB and SNe Ia data. The result is shown
in Figure~3. Again, we must caution that systematic errors in either
the SNIa data (Riess98) or the CMB could affect this
result.

Nevertheless, it is heartening to see that the combined constraint
favors a location in this parameter space which has not been ruled out
by other observations, though there may be mild conflict with
constraints on $\Omega_\Lambda$ from gravitational lensing (Falco, Kochanek,
\& Mu\~noz 1998). In fact, the region selected by the SNIa and CMB observations
is in concordance with inflation, large-scale structure measurements, 
and the ages of stars (Ostriker \& Steinhardt 1995; Krauss \& Turner 1995).
The combined constraint removes much of the high $\Omega_{\rm m}$,
high $\Omega_\Lambda$ region which was not ruled out by the SNIa
data alone, as well as much of the high $\Omega_{\rm m}$, low
$\Omega_\Lambda$ region allowed by the CMB data alone. The combined
constraint is consistent with a flat universe, as $\Omega_{\rm tot}=\Omega_{\rm m}+\Omega_\Lambda
= 0.94 \pm 0.26$ for MLCS and $1.00\pm 0.22$ for $\Delta m_{15}(B)$ ($1\sigma$ errors).
The enormous redshift difference between the CMB and the SNIa makes it
dangerous to
generalize this result beyond a cosmological constant model because of
the possible time-dependence on $\alpha_{\rm x}$. But for 
an equation of state fixed after recombination, the combined constraints continue
to be consistent with a flat geometry as long as $\alpha_{\rm x} \lesssim -0.6$. 
With better estimates of the
systematic errors in the SNIa data and new measurements of the CMB
anisotropy, these preliminary indications should quickly turn into very
strong constraints (Tegmark {\it et al.} 1998).

\clearpage
\begin{figure}[h]
\plotfiddle{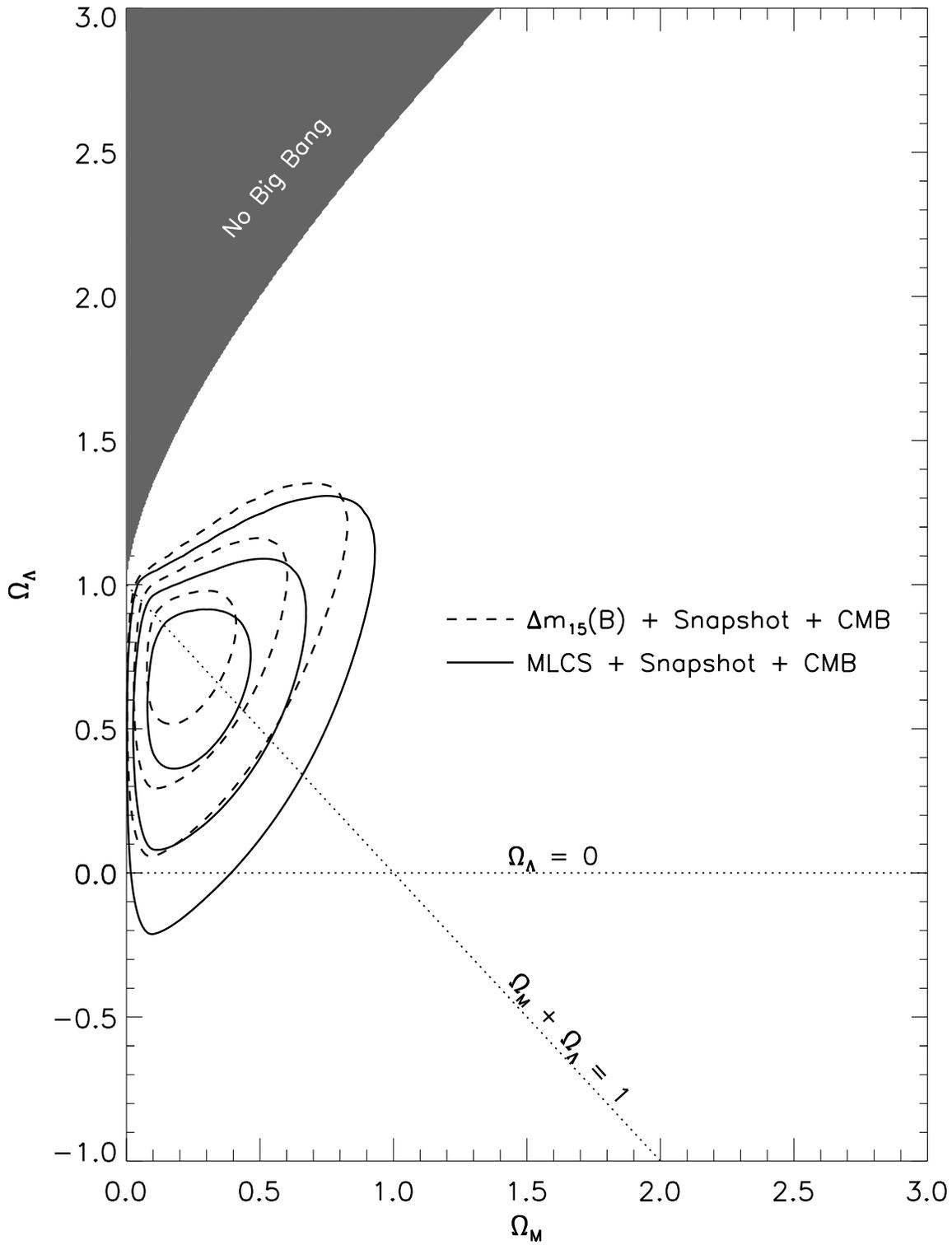}{11.0cm}{0}{80.0}{80.0}{-250.0}{-250.0}
\vspace*{8.0cm}
\caption{The combined constraints from SNIa
and the position of the first Doppler peak of the CMB angular power
spectrum. The equation of state parameter for the unknown component
is $\alpha_{\rm x}=-1$, like that for a cosmological constant. The contours
mark the 68\%, 95.4\%, and 99.7\% enclosed probability regions. }
\end{figure}

\section{Conclusions}

The current results from the High-Z Supernova Search Team
suggest that there is an
additional energy component sharing the Universe with gravitating matter.
For a flat geometry,
the ratio of the pressure of the unknown energy to its density
is probably more negative than $-0.6$. This effectively rules out topological defects
such as strings and textures as the additional component
and disfavors domain walls as that component. Open models are less
constrained, but favor $\alpha_{\rm x} <-0.5$. 
Although there are many intriguing
candidates for the ``X'' component, the current SNIa observations
imply that a vacuum energy
or a scalar field that resembles the cosmological
constant is the most likely culprit. 

Combining the SNIa probability
distribution with today's constraints from the position of the
first acoustic peak in the CMB power spectrum provides
a simultaneous observational
measurement of the densities of matter and of the unknown component.
Using CMB data from Hancock \etal\ (1998) and following the analysis
by White (1998)
the result favors a flat Universe, with $\Omega_{\rm tot}= 0.94 \pm 0.26$,
dominated by the ``X'' component
for $\alpha_{\rm x}\approx -1$.  Given the rapid improvement in
both the study of SN~Ia and the CMB, we can expect more powerful
inferences about the contents of the Universe to follow.

\acknowledgements

We thank U. Seljak and M. Tegmark
for some informative discussions and G. Rocha for
making the CMB likelihood function available before
publication. This work was
supported by grant GO-7505 from the Space Telescope
Science Institute, which is operated by the Association
of Universities for Research in Astronomy, Inc., under NASA
contract NAS5-26555, and at Harvard University
through NSF grants AST 92-21648 and AST 95-28899 and
an NSF Graduate Research Fellowship. Work at the
University of Washington was supported through NSF
grant AST-9617036.
A.V.F acknowledges the support of NSF grant AST-9417213.
S.M.C. was supported by NSF grant PHY/94-07194.

\end{document}